\documentstyle[epsfig]{mn}

\begin{document}

\title{Heating of the IGM by FRII radio sources}

\author[Christian R. Kaiser \& Paul Alexander]{Christian
R. Kaiser$^{1,2}$\thanks{present address: Max-Planck-Institut f{\"u}r
Astrophysik, Karl-Schwarzschild-Str. 1, 85740 Garching,
Germany. email: ckaiser@mpa-garching.mpg.de} and Paul Alexander$^1$\\
$^1$ MRAO, Cavendish Laboratory, Madingley Road, Cambridge, CB3 0HE,
UK\\ $^2$ University of Oxford, Department of Physics, Nuclear Physics
Laboratory, Keble Road, Oxford, OX1 3RH, UK}

\maketitle

\begin{abstract}
We present results of a numerical integration of the hydrodynamical
equations governing the self-similar, two-dimensional gas flow behind
the bow shock of an FRII radio source embedded in an IGM with a power
law density profile. The model predicts pressure gradients within the
cocoons consistent with modest backflow. For very steep external
density profiles sources may well not expand in a self-similar fashion
and in this case the model is not self-consistent. The assumption of
ram pressure confinement of the cocoons perpendicular to the jet axis
is found to overestimate the ratio of the pressure in front of the
radio hot spots and that in the cocoons. Based on the properties of
the gas between bow shock and cocoon we calculate the X-ray surface
brightness of the flow. This emission is found to be a good tracer of
the density distribution within the flow and varies significantly with
the properties of the unshocked IGM. The cooling-time of the shocked
IGM is found to be comparable to, or greater than, the Hubble
time. The influence of a radio source on the evolution of its gaseous
surroundings therefore extends well beyond the limited life time of
the source itself. We compare our results with the X-ray map of Cygnus
A and find some evidence for cold, dense gas clumps in the
surroundings of this object. The extended X-ray emission observed
around 3C 356 may also be caused by the bow shock of this radio
source. We also present an empirical model for the X-ray emission of
the shocked IGM due to thermal bremsstrahlung.
\end{abstract} 

\begin{keywords} hydrodynamics -- galaxies: active -- galaxies: individual: Cygnus A -- galaxies: individual: 3C 356 -- galaxies: jets -- X-rays: galaxies
\end{keywords}

\section{Introduction}

Most analytical models of FRII extragalactic radio sources are based
on the existence of two jets emerging from an AGN in the core of the
source; a model first proposed by Rees (1971)\nocite{mr71}. After
passing through a strong shock at the location of the radio hot spot,
the jet material inflates a cavity of hot, rarefied gas. The expansion
of this cocoon is supersonic with respect to the surrounding IGM and
therefore drives a bow shock into this material. This shock will
compress and heat the gas of the IGM, thereby increasing X-ray
emission from thermal bremsstrahlung. Inferred on theoretical grounds
by Scheuer (1974)\nocite{ps74}, this hot layer of gas between bow
shock and cocoon was first observed in Perseus A by B\"ohringer {\em
et al.}  (1993)\nocite{bvfen93} and in Cygnus A by Carilli {\em et
al.}  (1994)\nocite{cph94}. The extended X-ray emission detected in
the vicinity of 3C 356 (Crawford \& Fabian 1996\nocite{cf96}) may
represent another example of thermal bremsstrahlung of the IGM boosted
by the presence of a powerful radio source.

The first analytical investigation of a cavity expanding
supersonically into a surrounding atmosphere was presented by Sedov
(1959)\nocite{ls59} who considered the case of a strong explosion in a
uniform density profile. Dyson et al. (1980)\nocite{dfp80} extended
this analysis to spherically symmetric winds with a permanent energy
input and Heinz {\em et al.} (1998)\nocite{hrb98} included density
gradients by modeling the external density distribution with a King
(1972)\nocite{ik72} profile. The solutions of Sedov and Dyson {\em et
al.} are both self-similar. The expansion of both the cocoons and bow
shocks of FRII sources is shown to be self-similar regardless of its
geometrical shape by Kaiser \& Alexander (1997)\nocite{ka96b}
(hereafter KA), and this result suggests that a solution based on the
spherically symmetric case may be possible for a radio source provided
that the elongation of the cocoon along the axis of the jet can be
accommodated.  The solution for the flow of the shocked IGM between
bow shock and cocoon requires solution of at least a two-dimensional
problem instead of the one-dimensional spherical case of an explosion
or wind.

In this paper we assume the bow shock surrounding the cocoon of FRII
radio sources to be elliptical. We then determine the properties of
the shocked gas between bow shock and cocoon by integrating
numerically the relevant hydrodynamical equations. Analogous to KA, we
will assume that the density distribution of the IGM surrounding the
radio source is well described by a power law of the form $\rho _o
(r/a_o)^{-\beta}$, where $r$ is the radial distance from the centre of
the distribution.

Although we may choose the shape of the bow shock freely without
violating the conditions for self-similar expansion of the cocoon or
bow shock, we will show that another condition for self-similarity may
not be fulfilled by some of the solutions presented here. This
condition is the pressure equilibrium of the jets within the uniform
pressure of the cocoon away from the hot spots. This implies that for
these cases the model may not be self-consistent. One possibility, as
we discuss, is that the assumption of an elliptical shape for the bow
shock must be relaxed and the source expands self-similarly with a
somewhat modified shape. However, we believe that the results obtained
in this paper are an important first step in solving the problems
posed by the flow between bow shock and cocoon of FRII sources.

The general dynamics of the large-scale structure of FRII sources
depends on the details of the confinement of the cocoons of these
objects. It is usually assumed that the pressure of the material in
the cocoon is balanced by the ram pressure of the IGM (e.g. Scheuer
1974\nocite{ps74}, KA). In this paper we show that the assumption of
ram pressure confinement of the cocoons overestimates the ratio of the
pressure in front of the hot spot and that of the cocoon material,
leading to lower hot spot advance speeds in the analysis of KA.

The results of the integration of the hydrodynamical equations allows
us to determine the X-ray surface brightness of the shocked
material. The result of this calculation can be directly compared with
X-ray observations. The only currently available X-ray map of a
powerful FRII source with sufficient spatial resolution to allow such
a comparison, is the {\em ROSAT}\/ map of Cygnus A presented by
Carilli {\em et al.}  (1994)\nocite{cph94}. However, the advent of the
new X-ray telescopes such as {\em AXAF}\/ will allow us to extend this
analysis to a greater number of FRII sources.

In Section 2 we transform the hydrodynamical equations to an
elliptical coordinate system comoving with the self-similar expansion
of the bow shock and the cocoon. Section 3 describes the numerical
method used to integrate the equations. The properties of the flow are
discussed in Section 4. In Section 5 we calculate the X-ray emission
as predicted by the model and compare our results with the X-ray maps
of Cygnus A and 3C 356.

\section{The hydrodynamical equations in an elliptical coordinate system\label{sec:hydro}}

The hydrodynamical equations governing adiabatic fluid flow in
three dimensions if gravitational fields and viscosity can be
neglected, are the equation of motion:

\begin{equation} \frac{\partial}{\partial t} \,
{\bf v} + \left( {\bf v} \cdot {\bf \nabla} \right) {\bf v} = -
\frac{1}{\rho} {\bf \nabla} \, p,\label{eqmo}\end{equation}

\noindent the equation of continuity:

\begin{equation} \frac{\partial}{\partial t} \, \rho + {\bf \nabla} \cdot
\left( \rho {\bf v} \right) = 0\label{eqco}\end{equation}

\noindent and the adiabatic equation:

\begin{equation} \frac{\partial}{\partial t} \, \frac{p}{\rho ^{\Gamma}}
+ \left( {\bf v} \cdot {\bf \nabla} \right) \frac{p}{\rho ^{\Gamma}} =
0,\label{eqad}\end{equation}

\noindent where {\bf v} is the velocity vector of the gas, $p$ the
pressure, $\rho$ its density and $\Gamma$ its adiabatic index. We
assume in the following analysis that the bow shock surrounding the
cocoon is a prolate ellipsoid with rotational symmetry about the
$z$-axis defined by the jet, and therefore we rewrite these equations
in prolate spheroidal coordinates ($\eta$, $\xi$, $\phi = \pi /2$, see
Appendix \ref{sec:nonmov}). The rotational symmetry reduces the
problem to two spatial dimensions.

\begin{eqnarray} c \, c_r \
\frac{\partial}{\partial t} \, {\bf v}' + \left( {\bf v}' \cdot {\bf
\nabla}' \right) {\bf v}' & = & - \frac{1}{\rho '} {\bf \nabla}' p',\nonumber\\
c \, c_r \
\frac{\partial}{\partial t} \, \rho ' + {\bf \nabla}' \cdot \left( \rho '
{\bf v}' \right) & = & 0,\nonumber\\
c \, c_r \
\frac{\partial}{\partial t} \, \frac{p'}{\rho '^{\Gamma}} + \left(
{\bf v}' \cdot {\bf \nabla}' \right) \frac{p'}{\rho '^{\Gamma}} & = &
0,\label{prosph}
\end{eqnarray}

\noindent where all dashed quantities are functions of the elliptical
coordinates $\eta$ and $\xi$, $c$ is constant and $c_r = \left( \sinh
^2 \eta +\sin ^2 \xi \right) ^{1/2}$.

As the bow shock expands it is possible for all times to find a
stationary prolate spheroidal coordinate system characterized by a
constant $c$ in which the bow shock is described as a coordinate
surface with constant $\eta =\eta _b$. The aspect ratio of the cocoon,
i.e. the ratio of its length to its width, $R_{b}$, and the `length'
of the bow shock, i.e. the point at which the cut of the bow shock
with the $xz$-plane crosses the $z$-axis, determine the particular set
of values for $c$ and $\eta _b$. KA show that the expansion of the bow
shock and of the cocoon is self-similar; in this case the bow shock
can be described as a coordinate surface with $\eta =\eta _b$ for all
times. This is achieved by setting $c=c_o L_b$, where $c_o$ is a
constant and $L_b=f(t)$ is the `length' of the bow shock as described
above (see Appendix \ref{sec:mov}). It follows that

\begin{eqnarray}
c_o \, L_b \, c_r \left( \frac{\partial}{\partial t} {\bf v}' +
\dot{\eta} \, \frac{\partial}{\partial \eta} {\bf v}' + \dot{\xi} \,
\frac{\partial}{\partial \xi} {\bf v}' \right) & & \nonumber\\
+ \left( {\bf v}' \cdot
{\bf \nabla}' \right) {\bf v}' & = & -\frac{1}{\rho '} {\bf \nabla}'
p',\nonumber\\ 
c_o \, L_b \, c_r \left( \frac{\partial}{\partial t}
\rho ' +
\dot{\eta} \, \frac{\partial}{\partial \eta} \rho ' + \dot{\xi} \,
\frac{\partial}{\partial \xi} \rho ' \right) & & \nonumber\\
+ {\bf \nabla}' \cdot \left( \rho '
{\bf v}' \right) & = & 0,\nonumber\\
c_o \, L_b \, c_r \left( \frac{\partial}{\partial t} \frac{p'}{\rho
'^{\Gamma}} +
\dot{\eta} \, \frac{\partial}{\partial \eta} \frac{p'}{\rho
'^{\Gamma}} + \dot{\xi} \,
\frac{\partial}{\partial \xi} \frac{p'}{\rho
'^{\Gamma}} \right) & & \nonumber\\ 
+ \left(
{\bf v}' \cdot {\bf \nabla}' \right) \frac{p'}{\rho '^{\Gamma}} & = &
0.\label{comov}
\end{eqnarray}

Because of the self-similar evolution of both the bow shock and the
cocoon the flow between them must also be self-similar. This allows a
further simplification of equations (\ref{comov}) by introducing
dimensionless variables for the velocity, density and pressure
of the shocked IGM in which the dependencies on time (i.e. on $L_b$
and $\dot{L} _b$) and the spatial coordinates are separated,

\begin{eqnarray} 
{\bf v}' & = & \dot{L} _b \, \left[ U_{\eta} (\eta,\xi) \, {\bf \hat{e}} _{\eta} +
U_{\xi} (\eta,\xi) \, {\bf \hat{e}} _{\xi} \right], \nonumber\\
\rho ' & = & \rho _o \, \left( \frac{L_b}{a_o} \right) ^{-\beta} \
R(\eta,\xi), \nonumber\\ 
p' & = & \rho _o \, \left( \frac{L_b}{a_o} \right)
^{-\beta} \, \dot{L} _b ^2 \ P(\eta,\xi),\label{dimless}
\end{eqnarray}

\noindent where we model the external density distribution with a
power law, $\rho _o (r/a_o)^{-\beta}$ with $\rho _o$ the central
density, $a_o$ the core radius and $r$ the radial distance from the
centre of the distribution.

Using the comoving coordinate system described above and replacing the
flow variables with their dimensionless counterparts we find 

\begin{eqnarray}
c_o \, c_r \, \frac{\ddot{L}_b L_b}{\dot{L}_b^2} \, U_{\eta} +
b_{\eta} \, \frac{\partial}{\partial \eta} \, U_{\eta} + b_{\xi} \,
\frac{\partial}{\partial \xi} U_{\eta} & & \nonumber\\
+ \left( c_g \, U_{\eta} - c_e
\, U_{\xi} - 2 \, c_o \, c_r \, c_e \, c_g \right) U_{\xi} + \frac{1}{R} \,
\frac{\partial}{\partial \eta} P & = & 0,\nonumber\\[0.5cm]
c_o \, c_r \, \frac{\ddot{L}_b L_b}{\dot{L}_b^2} \, U_{\xi} + b_{\eta}
\, \frac{\partial}{\partial \eta} \, U_{\xi} + b_{\xi} \,
\frac{\partial}{\partial \xi} U_{\xi} & & \nonumber\\
- \left( c_g \, U_{\eta} - c_e
\, U_{\xi} - 2 \, c_o \, c_r \, c_e \, c_g \right) U_{\eta} + \frac{1}{R} \, \frac{\partial}{\partial \xi} P & = & 0,\nonumber\\[0.5cm]
R \, \frac{\partial}{\partial \eta} U_{\eta} + b_{\eta} \,
\frac{\partial}{\partial \eta} R + R \, \frac{\partial}{\partial \xi}
U_{\xi} + b_{\xi} \, \frac{\partial}{\partial \xi} R & & \nonumber\\ +
\left[ \left( c_e + \coth \eta \right) U_{\eta} + \left( c_g + \cot
\xi \right) U_{\xi} - c_o \, c_r \, \beta \right] R & = & 0,\nonumber\\[0.5cm]
-\frac{\Gamma \, P \, b_{\eta}}{R} \, \frac{\partial}{\partial \eta} R
+ b_{\eta} \, \frac{\partial}{\partial \eta} P -\frac{\Gamma \, P \,
b_{\xi}}{R} \, \frac{\partial}{\partial \xi} R & & \nonumber\\
+ b_{\xi} \,
\frac{\partial}{\partial \xi} P + c_o \, c_r \left[ \left(\Gamma
-1\right) \beta +2 \,\frac{\ddot{L}_b L_b}{\dot{L}_b^2} \right] P & =
& 0, \label{fourth}\end{eqnarray} 

\noindent where we have split the equation of motion into two
equations corresponding to components parallel to ${\bf \hat{e}}
_{\eta}$ and ${\bf \hat{e}} _{\xi}$ respectively, and

\begin{equation} b_{\eta} = U_{\eta} - c_o \, c_r \, c_e \ \ , \ \
b_{\xi} = U_{\xi} + c_o \, c_r \, c_g.\label{bdef}
\end{equation}

Furthermore, the ratio of the length of the bow shock, $L_b$, and the
length of the cocoon, $L_c$, must also be a constant. Using the result

\begin{equation} L_c = c_1 \, \left( \rho _o \, a_o^{\beta} \right)^{-\frac{1}{5-\beta}} \,
Q_o^{\frac{1}{5-\beta}} \, t^{\frac{3}{5-\beta}}
\label{linear}
\end{equation}

\noindent from KA, where $Q_o$ is the jet power and $c_1$ is a
dimensionless constant, $\ddot{L}_b L_b/\dot{L}_b^2$ reduces to
$(\beta -2)/3$, and the time dependence in equations (\ref{fourth})
can be eliminated.

\section{Numerical method\label{sec:nummet}}

The set of equations (\ref{fourth}) represent a system of first order
partial differential equations. Although there are no time derivatives
in this set the general form is that of an initial value problem
because the values of the variables are known at the surface of the
bow shock and the solution has to be propagated from this surface at
$\eta _b$ inwards to smaller $\eta$ in the direction of the contact
discontinuity delineating the cocoon.

The projection of the bow shock onto the $xz$-plane is an ellipse with
$\eta _b=$ constant and is described by

\begin{equation} r_b = \pm \sqrt{c_o \, \sinh ^2 \eta _b + \tanh ^2 \eta
_b \, l_b^2}, \label{delin}\end{equation} 

\noindent where $l_b$ is the dimensionless coordinate defined by
$l_b=z/L_b$. At the tip of the bow shock where $l_b=1$, $r_b =0$, and
for $l_b=1/2$ one finds $r_b=1/(2 \, R_b)$. These conditions yield

\begin{equation} c_o = \sqrt{1-\frac{1}{3 \,
R_b^2}} \ \ , \ \ \cosh \eta _b = \frac{1}{c_o}.  
\end{equation}

To derive the initial conditions at the bow shock we need to know the
velocity of the bow shock perpendicular to its surface,
$v_{\perp}$, and from the appendix of KA we find

\begin{equation}
v_{\perp} = \frac{v_x - \frac{\partial r_b}{\partial z} \,
v_z}{\sqrt{\left( \frac{\partial r_b}{\partial z} \right) ^2 +1}}.
\end{equation}

\noindent Since the bow shock expands self-similarly, $v_x= x/L_b$ and
$v_z = l_b$. By assuming that the bow shock is strong and that the
adiabatic index of the external atmosphere is 5/3 this implies for the
assumed elliptical shape of the shock surface

\begin{eqnarray}
U_{\eta} (\eta _b , \xi) & = & \frac{3}{4} \, \frac{\sinh \eta
_b}{c_r}, \nonumber\\
U_{\xi} (\eta _b , \xi) & = & 0\nonumber, \\
R (\eta _b , \xi) & = & 4 \, c_o^{-\beta} \, \left( \cos ^2 \xi +
\sinh ^2 \eta _b \right) ^{-\frac{\beta}{2}},\nonumber\\
P (\eta _b , \xi) & = & \frac{3}{4} \, c_o^{-\beta} \, \left( \cos ^2 \xi +
\sinh ^2 \eta _b \right) ^{-\frac{\beta}{2}} \, \frac{\sinh ^2 \eta
_b}{c_r^2}.
\end{eqnarray}

Given the symmetry of the problem it is sufficient to integrate the
equations only between $\xi =0$ and $\xi = \pi/2$ which corresponds to
the $z$ and $x$-axes respectively, and the following boundary
conditions must apply along these coordinate lines:

\begin{eqnarray}
U_{\xi} & = & 0, \nonumber\\
\frac{\partial}{\partial \xi} U_{\eta} & = & 0, \nonumber\\
\frac{\partial}{\partial \xi} R & = & 0, \nonumber\\
\frac{\partial}{\partial \xi} P & = & 0. 
\end{eqnarray}

The numerical integration was performed using a finite difference
discretisation scheme employing NAG\footnote{NAG is a registered
trademark of Numerical Algorithms Group Ltd.} routines on a grid of
1001 points between $\xi=0$ and $\xi=\pi/2$ separated by equal
intervals of length $\pi/2000$. Equations (\ref{fourth}) are
non-linear and it is therefore not straightforward to assess the
stability of the numerical method used. Trial calculations show that
in particular the solution for $U_{\xi}$ is subject to
oscillations. To suppress these oscillations, it is necessary to
introduce an `artificial viscosity' term of the form $a \, \partial ^2
U_{\xi}/ \partial \xi ^2$ to all equations in the system containing
expressions proportional to $\partial U_{\xi}/ \partial \xi$
(e.g. Richtmyer \& Morton 1967\nocite{rm67}). The constant $a$ is
chosen to be small compared to any other term in the system of
equations in order to keep the influence of the `artificial viscosity'
within the error of the integration method.

The integration has to be stopped at the contact discontinuity. There
can be no gas flow across this surface and therefore the criterion for
stopping the integration is met when the component of the gas velocity
perpendicular to the contact discontinuity is equal to the
corresponding component of the self-similar expansion velocity of this
surface. The shape of the contact surface, $r_c(z)$, and therefore the
self-similar expansion velocity, $z \, {\bf \hat{e}}_z + r_c (z) \,
{\bf \hat{e}}_x$, is not known {\em a priori}. Since the bow shock
will be closest to the cocoon at the hot spot, one expects the flow to
first reach the expansion velocity of the cocoon at $\xi=0$ and some
value $\eta = \eta_c$, i.e. on the $z$-axis. Assuming that $\partial
r_c / \partial z \rightarrow \infty$ at this point, the condition for
stopping the integration on the $z$-axis becomes $U_{\eta} (\eta _c ,
\xi =0) =c_o \, \cosh \eta _c$, where we have again made use of the
fact that the cocoon is expanding self-similarly. A comparison with
equation (\ref{bdef}) shows that at this point $b_{\eta} = 0$ which
means that the system of equations (\ref{fourth}) breaks down, as one
would expect at a contact discontinuity.

The integration was advanced from the bow shock at $\eta = \eta _b$ to
smaller values of $\eta$ in steps of $\eta _b / 5\times 10^4$ and
after each step it was checked whether the flow had reached the cocoon
at the hot spot, i.e. whether the velocity of the flow along the
$z$-axis was within 1\% of the self-similar velocity.

\setcounter{footnote}{1}

The cocoon is not elliptical and therefore does not coincide with the
coordinate surface at $\eta _c$. The integration was continued by
considering the grid point of the finite difference mesh of smallest
$\xi$ for which the solution had not yet reached the contact
surface. All derivatives with respect to $\xi$ in equations
(\ref{fourth}) were approximated by linear extrapolation using two
values of these expressions separated by $\eta _b /5\times 10^2$ along
a line of constant $\xi$. This extrapolation effectively converts
equations (\ref{fourth}) into a system of ordinary differential
equations which were solved using an implementation of the backward
differentiation formulae from the NAG{\footnotemark} library. The
integration was continued until the solution for the last point of the
grid had reached the cocoon.

\section{The gas flow between bow shock and cocoon\label{sec:flow}}

\begin{figure*}
\label{fig:sphecom}
\end{figure*}

An important test of the analysis and integration presented here is to
consider the case of a spherical symmetric wind as discussed by Dyson
{\em et al.} (1980)\nocite{dfp80}. These authors considered only the
case of a wind expanding into an uniform atmosphere, but using the
results of the previous section it is straightforward to extend their
analysis to more complex environments in which $\beta \ne 0$ (for a
solution using King density profiles see Heinz {\em et al.}
1998\nocite{hrb98}).

In the case of a spherical bow shock there is an additional degree of
rotational symmetry about the $x$-axis. This implies that one can set
$\xi =0$; $U_{\xi}$ and all derivatives with respect to $\xi$
vanish. Furthermore it is possible to change from the remaining
prolate spheroidal coordinate $\eta$ to the more familiar spherical
polar coordinate $r$ by using $r=c_o \cosh \eta$. Because of these
simplifications the second equation in (\ref{fourth}) vanishes
completely and we obtain

\begin{eqnarray}
\frac{\beta -2}{3} \, U_r + (U_r - r) \, \frac{\partial}{\partial r} U_r +
\frac{1}{R} \, \frac{\partial}{\partial r} P & = & 0,\nonumber\\
R \, \frac{\partial}{\partial r} U_r + (U_r - r) \, \frac{\partial}{\partial
r} R +2 \, \frac{U_r R}{r} - \beta R & = & 0,\nonumber\\
-\frac{\Gamma P}{R} \, (U_r - r) \, \frac{\partial}{\partial
r} R + (U_r - r) \, \frac{\partial}{\partial r} P & & \nonumber\\
+ \frac{3 \Gamma \beta
- \beta -4}{3} \, P & = & 0,
\end{eqnarray}

\noindent where we have used $R_b = 1/\sqrt{3}$ for a spherical bow
shock and therefore $\eta _b \rightarrow \infty$ which implies that
$\coth \eta \rightarrow 1$ for all $\eta$ close to $\eta _b$. Setting
$\beta =0$ recovers the equations used by Dyson {\em et al.}
(1980)\nocite{dfp80}. The initial conditions at the bow shock at $r=1$
for the spherical case are $U_r = 3/4$, $R=4$ and $P=3/4$.

Figure \ref{fig:sphecom} shows plots of the solution for the case of a
spherical bow shock. The curves for $\beta =0$ are identical to those
presented by Dyson et al. (1980)\nocite{dfp80}. In this case the
contact discontinuity is reached at $r_c=0.86$. For the two other
cases presented here we find $r_c=0.85$ and $r_c=0.84$ for $\beta =1$
and $\beta =2$ respectively. In all three cases the gas is accelerated
away from the cocoon but the acceleration is somewhat less effective
for the cases with a density gradient in the external atmosphere,
hence the greater detachment of the cocoon from the bow shock. While
this difference in the flow pattern is not very significant, the
distribution of the density and the pressure of the gas behind the bow
shock is distinctively different for each case. The gas is expanding
adiabatically after being compressed by the bow shock. This behaviour
is evident from the curves in Figure {\ref{fig:sphecom} for $\beta =0$
but is masked for $\beta = 2$ by the presence of dense gas at high
pressure close to the cocoon which has passed through the bow shock
earlier than the material in the regions closer to the bow shock. The
density and pressure of the external gas in front of the bow shock
decreases faster than that of the material which is expanding after
passing through the bow shock, giving the impression that the gas is
further compressed after passing through the bow shock. In this case
the cocoon is surrounded by a relatively thin, heavy shell of highly
compressed gas, while for $\beta =0$ the gas closest to the contact
discontinuity is rarefied.

Results from the numerical integrations of equations (\ref{fourth})
for the more complicated case of ellipsoidal bow shocks applicable to
FRII sources are summarized in Table \ref{tab:results}. Note, that the
stand-off distances between bow shock and cocoon are small both at
the hot spot and close to the $x$-axis. The shape of the cocoon is
therefore also close to an ellipsoid and this agrees well with
observations (e.g. Leahy \& Williams, 1984\nocite{lw84}). The
following sections describe the results in greater detail.

\begin{table*}
\label{tab:results}
\end{table*}

\subsection{The velocity field}

\begin{figure*}
\label{fig:velfield}
\end{figure*}

\begin{figure*}
\label{fig:velblow}
\end{figure*}

Figure \ref{fig:velfield} shows a cut defined by the $xz$-plane of the
velocity field of the gas flow between bow shock and contact
discontinuity for one half of a wide ($R_{b}=1$) ellipsoidal bow shock
in an uniform atmosphere. Because of the rotational symmetry about the
$z$-axis the vectors are lying fully in the $xz$-plane and there is a
reflection symmetry about the $z$-axis. The vectors show the velocity
of the flow with respect to the comoving coordinate system, with the
bow shock and cocoon at rest.

A closer examination of the flow close to the hot spot shows the flow
bending around the tip of the cocoon and then moving along the contact
discontinuity (Figure \ref{fig:velblow}, left). On the $z$-axis, the
axis of the jet, the flow ends in a stagnation point at the hot spot,
similar to the case of a spherical bow shock. Close to the $x$-axis
the flow is decelerated and there is a stagnation point on the
$x$-axis ahead of the cocoon (Figure \ref{fig:velblow}, right). This
behaviour is imposed on the solution by the assumed symmetry. The flow
pattern is very similar for other values of $\beta$ and $R_b$.

\subsection{Density and pressure}

\begin{figure*}
\label{fig:dengray}
\end{figure*}

The density distribution associated with the flow is shown in Figure
\ref{fig:dengray}. For a uniform atmosphere the density of the shocked
gas decreases towards the cocoon to a value which is below that of the
external density, similar to the behaviour of the gas behind a
spherical bow shock. For $\beta =1$ the density decreases slowly from
the bow shock to the cocoon, again similar to the spherical case
(Figure \ref{fig:sphecom}), however, the density of the gas just
behind the shock now decreases monotonically with increasing $x$
because of the external density gradient. For $\beta =2$ a similar
trend with increasing $x$ is seen, together with an increase in
density from the bow shock to the cocoon.

\begin{figure*}
\label{fig:pregray}
\end{figure*}

The pressure distribution of the gas between bow shock and contact
discontinuity is shown in Figure \ref{fig:pregray}. For a uniform
external density the pressure decreases monotonically from the hot
spot towards the $x$-axis, whereas for a decreasing density profile
the pressure firstly decreases away from the hot spot then increases
again towards the $x$-axis. For $\beta =2$ the pressure close to the
$x$-axis almost reaches the pressure in front of the hot spot.

In the analysis presented in KA which led to the prediction of
self-similar expansion of bow shock and cocoon we showed that it is a
good approximation to assume a constant pressure distribution within
the cocoon due to a high sound speed in this region. Across the
contact surface pressure is of course continuous. The results shown in
Figure \ref{fig:pregray} indicate a pressure variation around the
contact surface which is a minimum at some point between the hot spot
and the $x$-axis. In fact, within the cocoon there must be a backflow
(e.g. Norman {\em et al.} 1982\nocite{nsws82}) which is initially
accelerated from the hot spot and must, given the assumed symmetry, be
decelerated towards the symmetry plane of the $x$-axis. For most
values of $\beta$ this pressure variation is small and therefore we
believe the expansion of the source will remain approximately
self-similar and the results of KA and the analysis presented here
will apply. For $\beta$ close to 2 the pressure variation becomes
significant. The solution presented here will then be valid while the
source has an approximately elliptical shape, however in this case the
source is unlikely to expand self-similarly. This is because one of
the assumptions for self-similarity is the pressure equilibrium of the
jets within the radio cocoon the pressure within which is assume to be
uniform away from the hot spots. At least two possibilities for the
evolution of such a source exist. Firstly, if the source retains
approximate axial symmetry then as the high sound speed in the cocoon
leads to pressure equalization, the excess pressure in the shocked gas
will cause the cocoon to collapse resulting in a `pinching off' of the
cocoon; the bow shock will then not follow an elliptical coordinate
surface, although subsequent expansion may still be approximately
self-similar. Alternatively, the axial symmetry of the problem may not
be retained leading to significant off-axis flow in the cocoon as is
commonly observed.

Note also that with the steeper rise of the pressure in direction of
the core for large $\beta$ the density is increasing as well in these
regions. This implies that radiative cooling due to thermal
bremsstrahlung becomes important and our assumption of adiabatic
conditions may not be valid at this point. Any cooling will decrease
the pressure and including radiative processes in the analysis should
alleviate the problem of the steep pressure gradients towards the
core. However, including these effects is beyond the scope of this
paper and in the present form the existence of the pressure gradients
described above means that the model is not fully self-consistent for
large values of $\beta$.

\begin{figure*}
\label{fig:bigpan}
\end{figure*}

A comparison of the density and pressure distributions for different
aspect ratios of the bow shock but for the same external atmosphere is
shown in Figure \ref{fig:bigpan}. The gradients in the density and the
pressure distribution are greater for higher $R_b$. The variation in
pressure is again seen to be smaller than in density. The stand-off
distance between bow shock and cocoon at the hot spot, $\Delta _h$,
decreases for increasing $R_b$, while it stays almost constant for
constant $R_b$ but increasing $\beta$ (see Table
\ref{tab:results}). For large values of $R_b$ the stand-off distance
at the $x$-axis, $\Delta _x$, decreases with increasing $\beta$.

We conclude that for $\beta$ somewhat less than 2 self-similar
expansion can occur with a flow distribution and source shape well
approximated by the analysis presented here. For $\beta$ close to 2
the source may pass through a phase in which the analysis of the
present paper applies, but this does not represent a self-similar
expansion and deviations from the assumptions of axial symmetry or
elliptical bow shocks must occur.

\subsection{Confinement of the cocoon}

In our analysis presented in KA the ratio of the pressure at the hot
spot, $p_h$, and that of the cocoon, $p_c$, determines the value of
the dimensionless constant $c_1$ which describes the expansion of the
cocoon. In the case of a cylindrical cocoon which is confined by the
ram pressure of the receding IGM a simple estimate gives

\begin{equation}
\frac{p_h}{p_c} = 4 \, R^2.
\label{ram}
\end{equation}

This estimate ignores the variation in external density along the
outside of the cocoon and the small amount of backflow. Taking $p_c$
to be the pressure at the contact discontinuity half way between the
core and the hot spot, we find that our results are well fitted by
empirical relations of the form

\begin{equation}
\frac{p_h}{p_c} = \left\{
\begin{array}{ll} 
2.51 \, R^{1.95} & \mbox{; $\beta =0$}\\
1.95 \, R^{1.63} & \mbox{; $\beta =1$}\\
1.03 \, R^{1.57} & \mbox{; $\beta =2$}
\end{array} \right..
\label{ratio}
\end{equation}

\noindent The assumption of ram pressure confinement of the cocoon,
equation (\ref{ram}), therefore leads to an overestimate of the ratio
$p_h/p_c$. 

Because of the self-similar expansion of the cocoons, the cocoon
volume is given as $c_3 L_c^3$, where $L_c$ is the length of the
cocoon and $c_3$ is a constant depending on the cocoon geometry. The
values of $c_3$ derived from the analysis presented here are given in
Table \ref{tab:results}. For the assumption of a cylindrical cocoon
$c_3$ equals $\pi / (4 \, R^2)$; we find a maximal deviation of only
28\% of this approximation from our results.

The profile of the density distribution in the environments of
extragalactic radio sources is well fitted by a King
(1972)\nocite{ik72} profile. For distances from the centre of the
density distribution greater than the core radius, $a_o$, the King
profile is well fitted by the power law distribution, and for
distances smaller than $a_o$ a constant density $\rho _o$ is a good
approximation. Consider a radio source which starts out in such a
uniform density environment and expands beyond the core radius. KA
argue that the ratio $p_h / p_c$ is determined by intrinsic properties
of the jet since the cocoon and the jet are assumed to be in pressure
balance. In this case equations (\ref{ratio}) imply that the cocoon of
such a source will become narrower when it emerges from the uniform
density regime into the region with a decreasing density
gradient. This result is in agreement with the observation of Black
(1992)\nocite{ab92} who notes that in his sample shorter sources tend
to be `fatter'.

\section{X-ray emission from the heated IGM\label{sec:xray}}

\begin{figure*}
\label{fig:temgray}
\end{figure*}

Assuming the shocked IGM between bow shock and contact discontinuity
is an ideal gas consisting entirely of ionised hydrogen the absolute
temperature at any given point in the flow is

\begin{equation}
T = \frac{m_p}{k} \, \dot{L}_b^2 \, \frac{P(\eta , \xi)}{R(\eta ,
\xi)},
\label{temp}
\end{equation}

\noindent where $m_p$ is the mass of a proton and $k$ is the Boltzmann
constant. Note, that the temperature of the gas in the flow does not
depend on the temperature of the environment the source is expanding
into. 

Figure \ref{fig:temgray} shows a plot of the ratio $P/R$ for various
atmospheres. The differences between the temperature distribution for
different $\beta$ are relatively small. The thin shell of rarefied gas
around the cocoon in an uniform atmosphere is slightly hotter than its
surroundings and the overall temperature of the gas in the flow
decreases somewhat for increasing $\beta$. A comparison of models for
various aspect ratios (Figure \ref{fig:bigpan2}, left column) shows an
increasing prominence of the hot spot region with increasing $R_b$
which can be explained by the faster expansion speed of the bow shock
at the hot spot with respect to the expansion perpendicular to the jet
axis. Again the temperature scale is very similar in all four panels.
 
\begin{figure*}
\label{fig:bigpan2}
\end{figure*}

The specific emissivity of ionised hydrogen at temperature $T$ due to
thermal Bremsstrahlung is (e.g. Shu, 1991)\nocite{fs91}

\begin{equation} \epsilon _{\nu} = 7 \times 10^{-51} \frac{n_e \,
n_p}{\sqrt{T}} \, e^{-\frac{h \, \nu}{k \, T}} \ \mathrm{\frac{W}{m^3
\, Hz}},
\label{them}
\end{equation}

\noindent where $n_e$ and $n_p$ are the number densities of the
electrons and of the protons respectively. When observed with a finite
bandwidth, $\nu_1 \rightarrow \nu_2$, the integrated X-ray emissivity
is given by

\begin{eqnarray}
\lefteqn{\epsilon = 5.6\times 10^{11} \, \frac{\left( \rho _o \,
a_o^{\beta} \right) ^2}{a_1} \, L_b ^{\frac{-5\beta -2}{3}} \,
\sqrt{R^3 \, P} \, } \nonumber\\ 
& \times &\left[ e^{-4.0\times 10^{-7}
\, a_1^2 \, L_b^{\frac{4-2\beta}{3}} \, \nu \, \frac{R}{P}} \right]
_{\nu_2}^{\nu_1}\ \mathrm{\frac{W}{m^3}},
\label{emission}
\end{eqnarray}

\noindent with

\begin{equation}
a_1 \equiv \frac{5-\beta}{3} \, c_1^{\frac{\beta -5}{3}} \, \left( \rho
_o \, a_o^{\beta} \right) ^{\frac{1}{3}} \, Q_o^{-\frac{1}{3}},
\end{equation}

\noindent where we have used equation (\ref{linear}). To obtain the
X-ray surface brightness this expression has to be integrated along
the line of sight through the flow. The dependence of $\epsilon$ on
the advance speed of the bow shock, $\dot{L}_b$, is non-linear. To
proceed we therefore have to choose specific values for the external
density profile, the jet power, the linear size of the cocoon, $L_j$
and the observing frequency. Assuming $\rho _o = 5\times 10^{-22}$
kg/m$^3$, $a_o = 1$ kpc, typical values for the gas surrounding
isolated galaxies (Canizares {\em et al.}  1987\nocite{cft87}), $Q_o =
10^{39}$ W (Rawlings \& Saunders 1991\nocite{rs91}), $L_j=100$ kpc and
an observing band from $\nu _1 =0.1$ keV to $\nu _2 =2$ keV
(appropriate for the High Resolution Imager, {\em HRI}\/, of the {\em
ROSAT}\/ satellite) we obtain the results shown in Figure
\ref{fig:brigray}. Here we have assumed that the source is
rotationally symmetric about the $z$-axis, that the source lies
exactly in the plane of the sky and we have neglected the cosmological
redshift of the observing frequency.

\begin{figure*}
\label{fig:brigray}
\end{figure*}

The appearance of the flow in X-ray emission varies significantly with
$\beta$. For a uniform density environment the hot spot region
dominates the emission, while for higher values of $\beta$ the region
close to the $x$-axis is brightest and the emission is concentrated
towards the edge of the cocoon. Together with the observation made
earlier that the temperature variation is not very large within the
flow it is now evident that the X-ray surface brightness for the
frequency band chosen is a good tracer of the density
distribution. X-ray observations of FRII sources with high resolution
can therefore constrain the shape of the density distribution of the
atmosphere the source is expanding into. A similar result was found by
Clarke {\em et al.} (1997) from numerical simulations of a cocoon-bow
shock structure around a jet. For smaller aspect ratios the X-ray
surface brightness is more diffuse (see Figure \ref{fig:bigpan2},
right column), but the general features of the appearance of the
source remain the same. Note however the difference in the ages of the
sources of the same linear size but of different geometry and/or
environment.

Heinz {\em et al.} (1998)\nocite{hrb98} introduce observational
diagnostics for X-ray observations of extragalactic radio
sources. These are based on their assumption of spherical geometry for
the bow shock and cocoon which implies a uniform surface brightness at
all points within the flow at equal distance from the centre of the
source. From the analysis presented here it is clear that the X-ray
surface brightness of the flow between bow shock and cocoon can vary
considerably within the flow region if the source geometry is
elongated and/or the density gradient of the external medium is
steep. In these cases it is not straightforward to infer the absolute
values of source or environment parameters from X-ray observations
alone.

The total radiation due to thermal bremsstrahlung, $dE/dt$, can be
found by setting $\nu _1 =0$ and $\nu _2 \rightarrow \infty$ in
equation (\ref{emission}) and integrating the resulting emissivity
over the volume of the flow. For almost all the cases investigated
here we find the total luminosity to be significantly less than 10\%
of the jet power. We therefore conclude that our adiabatic treatment
of the flow between bow shock and cocoon is justified.

A rough estimate for the local cooling time of the shocked IGM can be
obtained by dividing the local energy density, $P(\eta,\xi)$, by
$dE/dt$. For all cases presented in this paper we find cooling times
of the order of 10\% to a few thousand times the Hubble time which is
much greater than the inferred life times of radio sources. The
presence of a radio source therefore influences the evolution of its
environment far beyond the limited life time of the source itself.

\subsection{Application of the model to Cygnus A and 3C 356}

The high resolution {\em ROSAT}\/ map of Carilli {\em et al.}
(1994)\nocite{cph94} has sufficient spatial resolution to allow a
comparison of source features observed in the radio at low frequency
with those observed in X-rays. For the density distribution of the
material surrounding Cygnus A we find $\rho _o =1.4 \times 10^{-22}$
kg/m$^3$, $a_o = 25.6$ kpc and $\beta = 1.4$ from the fitting of the
X-ray surface brightness distribution expected from a King profile
($\Omega =1$, $H_o =50$ km s$^{-1}$ Mpc$^{-1}$) to the data presented
by Carilli {\em et al.} (1994)\nocite{cph94}. Note, that this is
different from the fit obtained by these authors. Arnaud {\em et al.}
(1987)\nocite{ajfcnsm87} determine the temperature of the gas to be
$4\times 10^7$ K. Carilli {\em et al.} (1994)\nocite{cph94} also
present a 347 MHz VLA map of Cygnus A from which it is possible to
determine the linear size and the aspect ratio of the cocoon. Assuming
that Cygnus A is lying exactly in the plane of the sky we find $L_j
\approx 105$ kpc and $R \approx 2$; the difference in the lengths of
the two cocoons is roughly 15\%. The results of an integration of the
hydrodynamical equations describing the flow between bow shock and
cocoon for these parameters is given in Table \ref{tab:results}. The
aspect ratio of the bow shock of Cygnus A implied by this calculation
is 1.6.

Using the model of Kaiser, Dennett-Thorpe \& Alexander
(1997)\nocite{kda97a} for the radio emission of FRII sources we have
calculated the radio spectrum of Cygnus A for the given
parameters. The model is in good agreement with the observed radio
luminosities in the Gigahertz range presented by Baars {\em et al.}
(1977)\nocite{bgpw77} for a jet power of $Q_o=2\times 10^{39}$
W. Using the value of $p_h/p_c =4.1$ and the model of KA\nocite{ka96b}
we find an age of $2.9 \times 10^7$ years for Cygnus A. The present
hot spot advance speed implied by this is $9.7 \times 10^{-3}$ c. Both
results agree well with those of Carilli {\em et al.}
(1991)\nocite{cpdl91}.

After subtracting the expected X-ray emission due to thermal
bremsstrahlung of a King profile with the parameters quoted from the
map presented by Carilli {\em et al.}  (1994)\nocite{cph94}, the
excess or deficit of the emission due to the presence of the radio
source becomes visible. The most interesting feature for this analysis
are the emission enhancements to the sides and just outside the
cocoons because they must be caused by the shocked IGM between bow
shock and cocoon. For the eastern cocoon the peak in the residual
emission lies towards the core of the radio structure and the excess
measured after the King profile has been subtracted is 0.21 counts per
second per 0.5 arcsecond pixel with the {\em HRI}\/ in the band
between 0.1 keV and 2 keV. This implies an X-ray emission of the gas
in this region corresponding to 0.27 counts per second per pixel with
the {\em HRI}\/ because the measured count rate along a line of sight
through the cocoon region, which we assume to be free of X-ray emitting
gas, is 0.06 counts per second per pixel less than that expected from
the King profile. For the western cocoon the situation is less clear
and the emission enhancement may not be significant.

The surface brightness map for the case of Cygnus A is very similar to
the one shown on the bottom right of Figure
\ref{fig:brigray}. Qualitatively the predictions of the numerical
calculation presented here agree well with the observations. The
emission enhancements are at the correct positions, close to the core
of the source just outside the cocoon. However, a closer examination
of the temperature distribution reveals some severe problems with the
application of the results found in the previous section to the case
of Cygnus A. The value of $P/R$ at the point of highest surface
brightness is roughly 0.012. Together with the derived advance speed
of the hot spot we find a temperature of $1.9\times 10^6$ K at this
point. This is well below the temperature of the King profile assumed
for the gas which surrounds the bow shock. Indeed, even the model
temperature just behind the bow shock close to the core of the source
is below the temperature of the unshocked gas. This is caused by the
assumption of strong shock conditions along the entire bow shock which
may not be applicable to the case of Cygnus A. 

Using the temperature of the ambient gas the peak in the X-ray surface
brightness predicted by our model is $4\times 10^{-16}$ W m$^{-2}$ per
0.5 arcsecond pixel which corresponds to 0.012 counts per second per
pixel with the {\em HRI}\/. For the conversion we have used the
Internet version of PIMMS\footnote{PIMMS was programmed by K. Mukai at
the High Energy Astrophysics Science Archive Research Center of NASA,
available at {\em http://heasarc.gsfc.nasa.gov/Tools/w3pimms.html}\/}
with a column density for the galactic medium of $3 \times 10^{25}$
m$^{-2}$ towards Cygnus A (Clarke {\em et al.}
1997\nocite{chc97}). This discrepancy may be explained by the
existence of cold, dense clumps of gas embedded in the otherwise
smooth distribution of gas in galaxy clusters (Fabian {\em et al.}
1994\nocite{fcem94}). The X-ray emission of a volume of gas after the
bow shock has passed through it is proportional to $\rho ^2 M
\sqrt{T}$, where $\rho$ is the pre-shock density of the gas, $M$ the
Mach number of the shock within the gas and $T$ the temperature. For
ideal gas conditions the Mach number is proportional to $v/\sqrt{T}$,
where $v$ is the velocity of the shock with respect to the gas. The
ratio of the X-ray luminosity of a cold, dense cloud and that of hot
IGM with the same volume is then given by $\rho_c^2 v_c / (\rho _x^2
v_x)$ , where subscripts $c$ and $x$ are used for the cloud material
and the hot IGM respectively. The cold clouds are small and will
therefore quickly re-establish pressure equilibrium with the shocked
hot IGM surrounding it after the bow shock has passed through it
(e.g. McKee \& Cowie 1975\nocite{mc75}). We can therefore set the
velocity of the bow shock within the cloud material to $\sim
\sqrt{\rho _x / \rho _c} v_x$. With this we find that the ratio of the
X-ray luminosity of a cold cloud and that of the hot IGM occupying the
same volume is given by $(\rho_c / \rho _x)^{3/2}$. The peak in the
X-ray emission observed in the eastern lobe of Cygnus A can thus be
explained if the gas in this region is about 5 times denser than the
otherwise smoothly distributed IGM.

The striking symmetry of the two emission peaks on either side of the
base of the eastern cocoon is difficult to explain in this
scenario. Note however, that there is another peak in the X-ray
emission situated to the south-west of the core well inside the
western cocoon. This could be another clump of cold material with
similar properties to the two discussed above, which is observed in
projection and therefore appears to lie within the cocoon.

Crawford \& Fabian (1996)\nocite{cf96} report the detection of
extended X-ray emission in the vicinity of the FRII radio galaxy 3C
356. They argue that the emission is the signature of an extended IGM
in a cluster around 3C 356 which makes this radio galaxy the most
distant object (z=1.079) in which cluster gas has been detected
directly. The resolution of the X-ray map of this object is
insufficient for any quantitative study. However, the observed
extension is aligned with the northern cocoon of 3C356 which suggests
that the X-ray emission of the IGM surrounding the northern cocoon may
be enhanced by the compression and heating of this material by the bow
shock of the radio source, similar to the case of Cygnus A.

\subsection{An empirical model for the X-ray emission}

\begin{figure*}
\label{fig:empcom}
\end{figure*}

It is straightforward to calculate the total X-ray luminosity of the
shocked IGM between the bow shock and the cocoon for a given frequency
and bandwidth, $L_X$, predicted by the model for a given set of source
parameters by integrating over the surface area of the bow
shock. However, this requires the numerical calculation to be
performed for this particular set of parameters. In the following we
develop an analytical approximation for the X-ray luminosity using
empirical relations of source parameters derived from the numerical
calculations.

The total mass of the external gas swept up by the bow shock during
the life time of the radio source, $M$, is given by

\begin{equation}
M = \int \rho _o \left( \frac{r}{a_o} \right) ^{-\beta} \, dV_b = 4 \,
\pi \, \rho _o \, a_o^{\beta} L_b ^{3-\beta} \, I,
\end{equation}

\noindent where

\begin{equation}
I \equiv \int _0 ^{\pi /2} \int _0 ^{\eta _b} \sinh \eta \sin \xi \,
\frac{\sinh ^2 \eta + \sin ^2 \xi}{\left( \cos ^2 \xi + \sinh ^2 \eta
\right)} \, d\xi \, d\eta.
\end{equation}

If we assume this mass to be uniformly distributed over the volume
occupied by the flow, the density in this region is found to be
$\overline{\rho} = M/(V_b -V_c) \equiv M / (a_2 \, L_b^3)$, where
$V_b$ and $V_c$ are the volume enclosed by the bow shock and that of
the cocoon respectively. By definition $V_b = 2 \pi L_b ^3 \left(
c_o^2 \sinh ^2 \eta _b -1/3 \tanh ^2 \eta _b \right)$ and $V_c = c_3
L_b ^3$.

We define $\overline{T}$ as the average of the temperature of the gas
in the flow and since the temperature variations within the flow
material are small, local deviations from $\overline{T}$ will be small
as well. For an ideal gas $\overline{T} = m_p/k \, \dot{L}_b^2 \,
\overline{P/R}$.

With the two assumptions of uniform density and uniform temperature in
the flow we find from equation (\ref{them})

\begin{eqnarray}
\overline{L}_X & = & 5.2 \times 10^{13} \, \overline{\rho} \,
\sqrt{\overline{T}} \, \left( V_b -V_c \right) \, \left[ e^{-4.8 \times
10^{-11} \, \nu / \overline{T}} \right] _{\nu _2} ^{\nu _1}
\nonumber\\ 
& = & 9.1 \times 10^{13} \, \frac{I^2}{a_2} \, \frac{\left(
\rho _o \, a_o^{\beta} \right) ^2}{a_1} \, L_b^{\frac{7-5\beta}{3}} \,
\sqrt{\overline{P/R}} \nonumber\\
& \times & \left[ e^{-4.0 \times 10^{-7} \, a_1^2 \,
L_b^{\frac{4-2\beta}{3}} \, 1/\overline{P/R}} \right] _{\nu _2} ^{\nu
_2}.
\label{emp}
\end{eqnarray}

\noindent This expression depends on $c_1$, $\overline{P/R}$ and
$I/a_2$ which are all functions of the exponent of the density profile
of the unshocked IGM, $\beta$, and the geometry of the FRII source
defined by the aspect ratios $R$ or $R_b$. Empirical expressions for
these functions can be derived from the results of the numerical
calculation. We find

\begin{eqnarray}
\overline{P/R} & = & \left\{
\begin{array}{llll}
0.17 \, R^{-1.60} & \mbox{or} & 0.12 \, R_b^{-1.74} & \mbox{; $\beta
=0$}\\ 
0.13 \, R^{-1.68} & \mbox{or} & 0.088 \, R_b^{-1.71} & \mbox{; $\beta 
=1$}\\
0.095 \, R^{-1.67} & \mbox{or} & 0.068 \, R_b^{-1.64} & \mbox{; $\beta 
=2$}
\end{array}
\right. , \nonumber\\
I/a_2 & = & \left\{
\begin{array}{llll}
0.029 \, R^{-1.89} & \mbox{or} & 0.019 \, R_b^{-2.05} & \mbox{; $\beta
=0$}\\ 
0.12 \, R^{-1.00} & \mbox{or} & 0.093 \, R_b^{-1.01} & \mbox{; $\beta
=1$}\\ 
0.75 \, R^{0.50} & \mbox{or} & 0.83 \, R_b^{0.49} & \mbox{; $\beta
=2$}
\end{array}
\right. , \nonumber\\ 
c_1 & = & \left\{
\begin{array}{llll}
1.22 \, R^{0.49} & \mbox{or} & 1.35 \, R_b^{0.53} & \mbox{; $\beta
=0$}\\ 
1.02 \, R^{0.63} & \mbox{or} & 1.18 \, R_b^{0.64} & \mbox{; $\beta
=1$}\\ 
0.71 \, R^{0.88} & \mbox{or} & 0.85 \, R_b^{0.86} & \mbox{; $\beta
=2$}
\end{array}
\right.
\end{eqnarray}

Figure \ref{fig:empcom} shows a comparison of equation (\ref{emp})
using the empirical relations above, with the results of the numerical
calculation for $\rho _o = 5\times 10^{-22}$ kg m$^{-3}$, $a_o = 1$
kpc, $Q_o = 10^{39}$ W, $\nu _1 =0.1$ kev and $\nu _2 = 2$ keV. For
$\beta > 0$ the empirical model agrees well with the numerical
results. The deviation for large linear sizes and $\beta =0$ is
acceptable in many cases since it is unlikely that a uniform density
region extends to very large distance from the centre of any object
hosting a radio source. Note however, that the fit of the empirical
model worsens for higher observing frequencies since the exponential
part of equation (\ref{emission}), which is not well fitted by the
empirical formula, then dominates this expression.

\section{Conclusions}

We presented a numerical integration of the hydrodynamical equations
governing the self-similar gas flow behind a strong shock in two
dimensions assuming the shock to have an ellipsoidal geometry. The
density distribution in the unshocked environment of the shock is
modeled with a power law. Applying this model to the bow shocks
surrounding the cocoons of extragalactic radio sources we found that
there should be a pressure gradient within the cocoons of sources
located in non-uniform environments which is consistent with the
existence of backflow within the cocoon. In extreme density profiles
non-axially symmetric flow or pinching-off of the cocoon is very
likely to occur. Furthermore, the expansion of the cocoon and bow
shock may not be self-similar. In these cases ($\beta \sim 2$) the
model is not self-consistent. We also find that the assumption of ram
pressure confinement of the cocoon perpendicular to the jet axis leads
to an overestimate of the ratio of the pressure in front of the hot
spot and the pressure in the cocoon.

From the properties of the gas in the flow we calculated the X-ray
surface brightness predicted by the model. Because of the small
variations of the temperature in the flow, the surface brightness was
found to be a good tracer of the gas density. The appearance of radio
sources in X-rays is found to vary significantly with the properties
of the external medium. Simple diagnostic tools have to be refined to
account for the elongation of the bow shock of most sources. The
cooling times of the IGM shocked by the bow shock are much longer than
the expected life time of the radio sources causing the bow shock. The
evolution of any concentration of matter in the universe will
therefore be significantly influenced by the presence of a radio
source in its centre. This is the case even if the activity time scale
of the source is much shorter than the evolution time scale of its
surroundings.

We compared the predictions of our model with the X-ray map of Cygnus
A of Carilli {\em et al.} (1994)\nocite{cph94}. Although our model
predicts the strongest X-ray emission at the position where it is
observed, on the outside of the cocoon closest to the core of the
source, the model can not reproduce the observed luminosity. The
emission may be caused by the presence of overdense clumps of gas
embedded in the otherwise smooth IGM.

The enhanced X-ray emission of the shocked IGM behind the bow shock of
a strong radio source may also explain the observed extended X-ray
emission around 3C 356.

Based on the numerical calculation we also presented an empirical
model for the total X-ray emission expected from the shocked IGM
surrounding the cocoons of powerful FRII sources.

Despite the inconsistencies within the model that arise for large
values of $\beta$ we believe that this analysis is an important first
step in determining the influence of powerful radio sources on the
properties and evolution of their gaseous environments.

\section*{Acknowledgments}

The authors thank S. Rawlings for drawing their attention to the X-ray
map of 3C 356 and the anonymous referee for her/his very helpful
comments.

\bibliography{../../crk} 
\bibliographystyle{../../mnras}

\setcounter{section}{1}
\setcounter{subsection}{0}
\setcounter{subsubsection}{0}
\renewcommand{\thesection}{\Alph{section}}

\section*{Appendix}

\subsection{Comoving prolate spheroidal coordinates\label{sec:ell}}

\subsubsection{Fixed coordinates\label{sec:nonmov}}

\begin{figure}
\label{fig:coord}
\end{figure}
 
The Cartesian coordinates of any point in space can be related to
their prolate spheroidal counterparts via the expressions

\begin{eqnarray} z & = & c \cosh \eta \, \cos \xi, \nonumber\\
x & = & c \, \sinh \eta \, \sin \xi \, \sin \phi, \nonumber\\
y & = & c \, \sinh \eta \, \sin \xi \, \cos \phi, \label{conver}
\end{eqnarray}

\noindent where $c$ is a constant describing the eccentricity of the
ellipsoids of constant $\eta$ (see Figure \ref{fig:coord}). Since we
assume that the bow shock is of ellipsoidal shape with the $z$-axis,
the jet axis, being the axis of rotational symmetry the problem will
be independent of $\phi$ and we seek a solution (without loss of
generality) in the $xz$-plane, i.e. $\phi = \pi /2$. The unit vectors
with respect to $\eta$ and $\xi$ are

\begin{eqnarray}
{\bf \hat{e}} _{\eta} & = & \frac{1}{c_r}
\, \left( \sinh \eta \, \cos \xi \ {\bf \hat{e}} _z + \cosh \eta \, \sin
\xi \ {\bf \hat{e}} _x \right),\nonumber\\
{\bf \hat{e}} _{\xi} & = & \frac{1}{c_r} \, \left( -\cosh \eta \, \sin
\xi \ {\bf \hat{e}} _z + \sinh \eta
\, \cos \xi \ {\bf \hat{e}} _x \right),
\end{eqnarray}

\noindent with

\begin{equation} c_r = \sqrt{\sinh ^2 \eta + \sin ^2 \xi} \end{equation}

\noindent It immediately follows that

\begin{eqnarray}
\frac{\partial}{\partial \eta} \, {\bf \hat{e}} _{\eta} = -c_g \, {\bf
\hat{e}} _{\xi} \ \ , \ \ \frac{\partial}{\partial \xi} \, {\bf \hat{e}}
_{\eta} = c_e \, {\bf \hat{e}} _{\xi}\nonumber\\
\frac{\partial}{\partial \xi} \, {\bf \hat{e}} _{\xi} = - c_e \, {\bf
\hat{e}} _{\eta} \ \ , \ \ \frac{\partial}{\partial \eta} \, {\bf \hat{e}}
_{\xi} = c_g \, {\bf \hat{e}} _{\eta}
\end{eqnarray}

\noindent where

\begin{eqnarray} c_e & = & \frac{\sinh \eta \, \cosh \eta}{c_r^2} \nonumber\\
c_g & = & \frac{\sin \xi \, \cos \xi}{c_r^2}. 
\end{eqnarray}

Using equations (\ref{conver}) the $\eta$ and $\xi$ components of the
${\bf \nabla}$-operator in this coordinate system can be found.

\begin{eqnarray} {\bf \nabla} _{\eta} & = & \frac{1}{c \, c_r} \,
\frac{\partial}{\partial \eta} \ \equiv \ \frac{1}{c \, c_r} \, {\bf \nabla}'_{\eta}\nonumber \\
{\bf \nabla} _{\xi} & = & \frac{1}{c \, c_r} \, \frac{\partial}{\partial \xi} \ \equiv \ \frac{1}{c \, c_r} \, {\bf \nabla}'_{\xi}
\end{eqnarray}

\noindent The divergence of a vector, ${\bf \nabla} \cdot {\bf V}$, in
prolate spheroidal coordinates assuming that ${\bf V}$ does not depend
on $\phi$ can be written in the form

\begin{eqnarray} c \, c_r \, {\bf \nabla} \cdot {\bf V} & = &
\frac{\partial}{\partial
\eta} {\bf V}_{\eta} + \left( c_e + \coth \eta \right) {\bf V}_{\eta}
 \nonumber\\
& & + \frac{\partial}{\partial \xi} {\bf V}_{\xi} + \left( c_g + \cot \xi
\right) {\bf V}_{\xi}. 
\end{eqnarray}

\subsubsection{Comoving coordinates\label{sec:mov}}

By setting $c=c_o L_b$ it is possible to transform the stationary
coordinate system discussed in the previous section into comoving
coordinates. From equations (\ref{conver}) it is clear that if $c$ is
a function of time, $\eta$ and $\xi$ must also depend on time. Taking
the derivative of (\ref{conver}) with respect to $t$ one finds

\begin{eqnarray}
\dot{\eta} & = & - \frac{\dot{L}_b}{L_b} \, c_e,\nonumber\\
\dot{\xi} & = & \frac{\dot{L}_b}{L_b} \, c_g,
\end{eqnarray}

\noindent where a dot denotes a time derivative. This also means that
all partial derivatives with respect to time in equations
(\ref{prosph}) are replaced according to

\begin{equation} \frac{\partial}{\partial t} \rightarrow
\frac{\partial}{\partial t} + \dot{\eta} \, \frac{\partial}{\partial
\eta} + \dot{\xi} \, \frac{\partial}{\partial \xi}.
\end{equation}

\noindent Note, that the partial derivatives with respect to the
spatial coordinates are unchanged and we therefore continue to use the
symbols, $\eta$ and $\xi$, for these spatial coordinates.

\setcounter{table}{0}

\begin{table*}
\caption{\footnotesize The results of the numerical
integration. $\beta$ is the exponent of the power law for the external
density distribution, $R_b$ is the aspect ratio of the bow shock, $R$
is the aspect ratio of the cocoon, $p_h/p_c$ is the ratio of the
pressure at the hot spot and the pressure in the cocoon defined as the
pressure at the contact discontinuity half way between the core of the
radio source and the tip of the cocoon, $\eta _b$ is the elliptical
coordinate surface coinciding with the bow shock, $\eta _h$ is the
elliptical coordinate at which the flow reaches the cocoon in front of
the hot spot, $\eta _c$ is the elliptical coordinate at which the flow
reaches the point on the contact discontinuity half way between the
core and the tip of the cocoon, $\eta _x$ is the elliptical coordinate
at which the flow reaches the cocoon on the $x$-axis, $\Delta _h$ is
the stand-off distance of the bow shock from the contact discontinuity
at the hot spot, $\Delta _x$ is the stand-off distance of the bow
shock from the cocoon at the $x$-axis and $c_3$ is the volume
constant.}  
{\centering
\begin{tabular}{llccccccccc}
& $R_b$ & $R$ & $p_h/p_c$ & $\eta _b$ & $\eta _h$ & $\eta _c$ & $\eta _x$ &
$\Delta _h$ & $\Delta _x$ & $c_3$\\ 
\hline 
& $1.0$ & 1.221 & 3.518
& 0.6585 & 0.5775 & 0.5386 & 0.5373 & 0.0435 & 0.1160 & 0.4153\\ 
& $2.0$ & 2.679 & 16.92 & 0.2971 & 0.2581 & 0.2217 & 0.2217 & 0.0105
& 0.0741 & 0.0940\\
$\beta = 0$ & $3.0$ & 4.125 & 39.80 & 0.1949 & 0.1699 & 0.1415 
& 0.1415 & 0.0045 & 0.0528 & 0.0403\\
& $4.0$ & 5.559 & 71.63 & 0.1454 & 0.1268 & 0.1044 & 0.1044 & 0.0025 
& 0.0406 & 0.0223\\
& $5.0$ & 6.985 & 111.8 & 0.1160 & 0.1022 & 0.0828 & 0.0828 & 0.0015 
& 0.0329 & 0.0146\\[2ex]
& $1.0$ & 1.217 & 2.384 & 0.6585 & 0.5715 & 0.5386 & 0.5360 & 0.0465 
& 0.1172 & 0.4153\\
& $2.0$ & 2.579 & 7.973 & 0.2971 & 0.2581 & 0.2300 & 0.2276 & 0.0105
& 0.0683 & 0.1015\\
$\beta = 1$ & $3.0$ & 3.858 & 16.67 & 0.1949 & 0.1699 & 0.1512 & 
0.1477 & 0.0045 & 0.0466 & 0.0461\\
& $4.0$ & 4.997 & 28.34 & 0.1454 & 0.1268 & 0.1160 & 0.1142 & 0.0025
& 0.0308 & 0.0276\\
& $5.0$ & 6.297 & 43.34 & 0.1160 & 0.1022 & 0.0919 & 0.0919 & 0.0015
& 0.0239 & 0.0174\\[2ex]
& $1.0$ & 1.211 & 1.509 & 0.6585 & 0.5674 & 0.5400 & 0.5321 & 0.0485
& 0.1209 & 0.4178\\
& $2.0$ & 2.451 & 3.704 & 0.2971 & 0.2541 & 0.2413 & 0.2264 & 0.0115
& 0.0695 & 0.1122\\
$\beta = 2.0$ & $3.0$ & 3.611 & 7.170 & 0.1949 & 0.1699 & 0.1614 &
0.1473 & 0.0045 & 0.0470 & 0.0527\\
& $4.0$ & 4.758 & 11.96 & 0.1454 & 0.1268 & 0.1218 & 0.1218 & 0.0025
& 0.0232 & 0.0305\\
& $5.0$ & 5.895 & 17.99 & 0.1160 & 0.1022 & 0.0981 & 0.0981 & 0.0015
& 0.0176 & 0.0199\\
\hline
\multicolumn{11}{c}{Cygnus A}\\[1ex]
$\beta =1.4$ & 1.6 & 2.031 & 4.099 & 0.3779 & 0.3256 & 0.2970 & 0.2917 &
0.0175 & 0.0842 & 0.1610\\
\hline
\end{tabular}}
\end{table*}

\begin{onecolumn}

{\bf Figure 1:} The flow behind a spherical bow shock in various
atmospheres. The top panel shows the velocity of the flow, the bottom
panels the density and the pressure. Solid curves: $\beta =0$, dashed
curves: $\beta =1$ and dotted curves: $\beta =2$.\\[2ex]

{\bf Figure 2:} The velocity field of the flow between bow shock and
contact discontinuity in the frame of the comoving coordinate system
for $\beta =0$ and $R_{b} =1$. The bow shock defines the upper edge,
the contact discontinuity the lower edge of the field plotted. The AGN
which drives the jet is located in this plot at the point ($z=0$,
$x=0$) and the tip of the bow shock is at ($z=0$, $x=1$).\\[2ex]

{\bf Figure 3:} Details of the flow pattern shown in Figure
\ref{fig:velfield}. Left: the flow close to the hot spot. Right: the
flow near the $x$-axis. Note, that the vectors representing the flow
velocity have the same length scale in both plots.\\[2ex]

{\bf Figure 4:} Density profile of the flow between bow shock and
contact discontinuity. Top: Uniform atmosphere, $\beta =0$. Bottom
left: $\beta =1$. Bottom right: $\beta =2$. For all plots $R_b=3$. The
gray scales are normalised such that the density just ahead of the tip
of the bow shock is the same in each panel.\\[2ex]

{\bf Figure 5:} Pressure profile of the flow between bow shock and
contact discontinuity. Top: Uniform atmosphere, $\beta =0$. Bottom
left: $\beta =1$. Bottom right: $\beta =2$. For all plots $R_b
=3$.\\[2ex]

{\bf Figure 6:} Comparison of the density and pressure distribution
for various aspect ratios of the bow shock. Left column: density,
right column: pressure. From top to bottom: $R_b =1$, $R_b =2$, $R_b
=4$ and $R_b =5$. For all plots $\beta =1$.\\[2ex]

{\bf Figure 7:} Temperature profile of the flow between bow shock and
contact discontinuity. Top: Uniform atmosphere, $\beta =0$. Bottom
left: $\beta =1$. Bottom right: $\beta =2$. For all plots
$R_b=3$.\\[2ex]

{\bf Figure 8:} Comparison of the temperature and X-ray surface
brightness distribution for various aspect ratios of the bow
shock. Left column: temperature, right column: surface
brightness. From top to bottom: $R_b =1$, $R_b =2$, $R_b =4$ and $R_b
=5$. For all plots $\beta =1$. For the chosen parameters (see text)
the ages of the radio sources in the right column are $3.0 \times
10^7$ years, $1.5 \times 10^7$ years, $8.8 \times 10^6$ years and $7.4
\times 10^6$ years respectively from top to bottom.\\[2ex]

{\bf Figure 9:} Surface brightness profile of the flow between bow
shock and contact discontinuity. Top: Uniform atmosphere, $\beta
=0$. Bottom left: $\beta =1$. Bottom right: $\beta =2$. For all plots
$R_b=3$. See text for the parameters of the external atmosphere and
the jet power. The resulting ages of the radio sources are: Top: $3.7
\times 10^7$ years, bottom left: $1.1 \times 10^7$ years and bottom
right: $3.5 \times 10^6$ years.\\[2ex]

{\bf Figure 10:} Comparison of the empirical model for the X-ray
emission with the numerical calculation. Individual symbols are
results from the numerical calculation while curves represent the
empirical model. Top panel: $R_b =1.0$, bottom left: $R_b =3$ and
bottom right: $R_b =5$. Solid lines and circles: $\beta =0$, dashed
lines and crosses: $\beta =1$ and dotted lines and squares: $\beta
=2$. See text for other model parameter.\\[2ex]

{\bf Figure 11:} Prolate spheroidal coordinates for $c=0.6$ and
rotation about the $z$-axis.

\end{onecolumn}

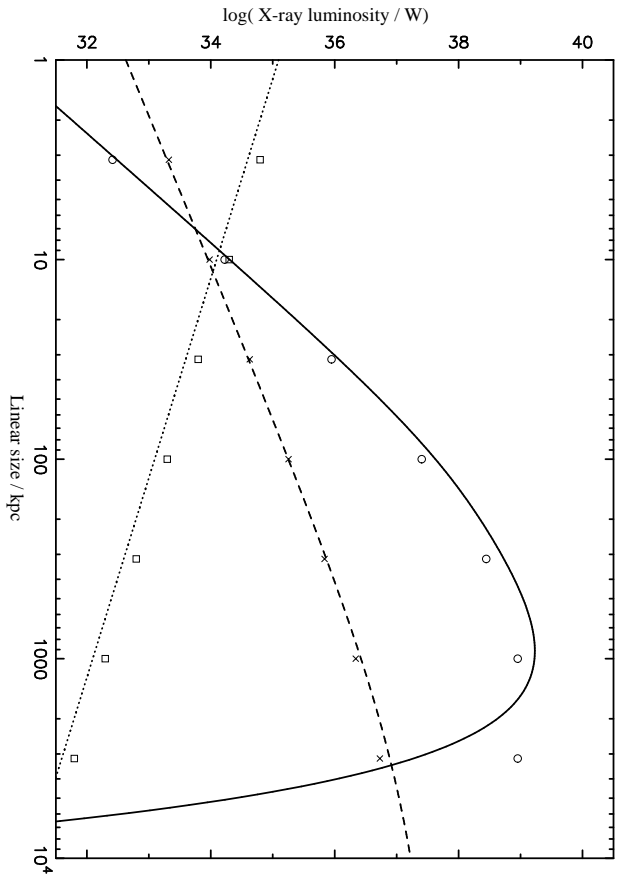
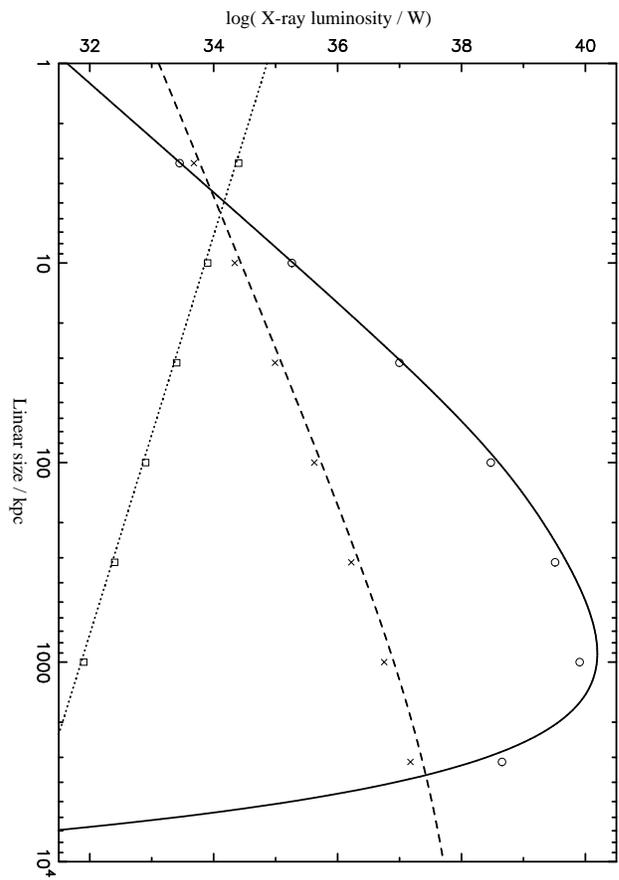
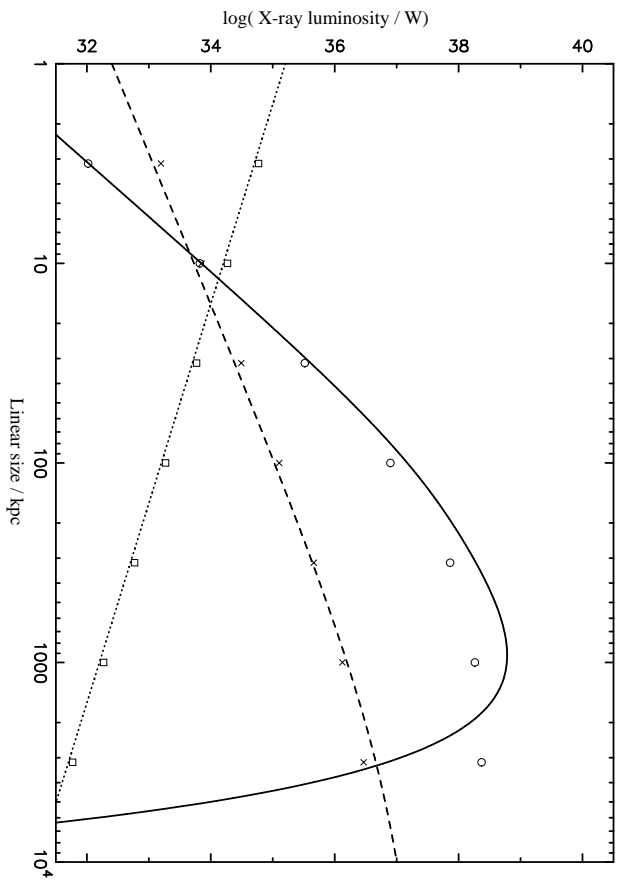
\end{document}